\documentclass[%
 reprint,
 amsmath,amssymb,
 aps,
]{revtex4-1}
\usepackage{graphicx}
\usepackage{dcolumn}
\usepackage{booktabs}
\usepackage{array}
\usepackage{bm}
 \usepackage{indentfirst}
 \usepackage{color}

\begin{document}


\title{An arbitrary two-particle high-dimensional Bell state measurement by auxiliary entanglement}%
\author{Hao Zhang}\email{These authors contribute equally to this work}
\affiliation{CAS Key Laboratory of Quantum Information, University of Science and Technology of China, Hefei, 230026, People's Republic of China}
\affiliation{CAS Center For Excellence in Quantum Information and Quantum Physics, University of Science and Technology of China, Hefei, 230026, People's Republic of China}

\author{Chao Zhang}\email{These authors contribute equally to this work}
\affiliation{CAS Key Laboratory of Quantum Information, University of Science and Technology of China, Hefei, 230026, People's Republic of China}
\affiliation{CAS Center For Excellence in Quantum Information and Quantum Physics, University of Science and Technology of China, Hefei, 230026, People's Republic of China}

\author{Xiao-Min Hu}
\email{huxm@ustc.edu.cn}
\affiliation{CAS Key Laboratory of Quantum Information, University of Science and Technology of China, Hefei, 230026, People's Republic of China}
\affiliation{CAS Center For Excellence in Quantum Information and Quantum Physics, University of Science and Technology of China, Hefei, 230026, People's Republic of China}

\author{Bi-Heng Liu}
\email{bhliu@ustc.edu.cn}
\affiliation{CAS Key Laboratory of Quantum Information, University of Science and Technology of China, Hefei, 230026, People's Republic of China}
\affiliation{CAS Center For Excellence in Quantum Information and Quantum Physics, University of Science and Technology of China, Hefei, 230026, People's Republic of China}

\author{Yun-Feng Huang}
\affiliation{CAS Key Laboratory of Quantum Information, University of Science and Technology of China, Hefei, 230026, People's Republic of China}
\affiliation{CAS Center For Excellence in Quantum Information and Quantum Physics, University of Science and Technology of China, Hefei, 230026, People's Republic of China}
\author{Chuan-Feng Li}
\email{cfli@ustc.edu.cn}
\affiliation{CAS Key Laboratory of Quantum Information, University of Science and Technology of China, Hefei, 230026, People's Republic of China}
\affiliation{CAS Center For Excellence in Quantum Information and Quantum Physics, University of Science and Technology of China, Hefei, 230026, People's Republic of China}
\author{Guang-Can Guo}
\affiliation{CAS Key Laboratory of Quantum Information, University of Science and Technology of China, Hefei, 230026, People's Republic of China}
\affiliation{CAS Center For Excellence in Quantum Information and Quantum Physics, University of Science and Technology of China, Hefei, 230026, People's Republic of China}

\date{\today}

\begin{abstract}

Bell state measurement (BSM) plays a vital role in quantum information. There are many researches on BSM of qubit Bell state, however, there is no definite solution of how to realize high-dimensional Bell state measurement (HDBSM). In this paper, We present a scheme for realizing arbitrary high-dimensional two-particle Bell state measurement by auxiliary entanglement. In our scheme, the maximal entangled states with different degrees of freedom (DOF) are used as auxiliary states, and then the HDBSM can be achieved by implementing Bell measurement with different degrees of freedom. We present the detailed HDBSM scheme for three and four-dimensional Bell state, and the general formula for arbitrary high-dimensional Bell state. Furthermore, We have designed the experimental scheme for three-dimensional BSM, and given the method of arbitrary dimension. This method can promote the high-dimensional quantum information task.
\end{abstract}

\maketitle
BSM as a crucial measurement in quantum mechanics, is widely used in quantum information tasks, such as quantum teleportation \cite{teleportation1,teleportation2,teleportation3}, superdense coding \cite{coding1,coding2,coding3}, quantum computing \cite{computing1,computing2,computing3}, etc. BSM has been implemented in many systems \cite{BSM1,BSM2}, especially in linear optics. However, a complete BSM can not be achieved with linear optics \cite{BSMN1,BSMN2} with current technology. For many years, researchers have been working for achieve a complete BSM with linear optics. For example, two-dimensional complete BSM can be achieved by local measurement using OAM as an auxiliary entanglement \cite{two-dimension1}. Or one can use time bin as an auxiliary DOF to complete BSM \cite{two-dimension2}.

In the field of quantum information, high-dimensional entanglement has many advantages over two-dimensional entanglement, such as higher information capacity, enhanced robustness against eavesdropping and quantum cloning \cite{clone1,clone2}, larger violation of local-realistic theories and its advantages in quantum communication \cite{large1,large2}. Therefore, the development of high-dimensional entanglement has become a hot research in quantum information. HDBSM, as the most important tool for studying high-dimensional entanglement, has also attracted extensive attention. However, there has no definite method to the implementation of HDBSM with linear optics.

In this paper, we present a method for completing arbitrary HDBSM via auxiliary entanglement. HDBSM of system state can be accomplished by BSM of different DOFs on single-particle, which has a great advantage, because in linear optics, it is much easier to complete a BSM with two degrees of freedom on a photon. We derive the three and four-dimensional entanglement BSM scheme, and extend the method to arbitrary HDBSM. Furthermore, we designed an experimental scheme for three-dimensional BSM.

We define the three-dimensional system state $\psi_{ij}^{\overline{d}}$ as
\begin{equation*}
 \psi_{00}^{\overline{3}}=(|00\rangle+|11\rangle+|22\rangle)/\sqrt{3},\\
\end{equation*}
\begin{equation*}
 \psi_{10}^{\overline{3}}=(|00\rangle+e^{2\pi i/3}|11\rangle+e^{4\pi i/3}|22\rangle)/\sqrt{3},\\
\end{equation*}
\begin{equation*}
 \psi_{20}^{\overline{3}}=(|00\rangle+e^{4\pi i/3}|11\rangle+e^{2\pi i/3}|22\rangle)/\sqrt{3},\\
\end{equation*}
\begin{equation*}
 \psi_{01}^{\overline{3}}=(|01\rangle+|12\rangle+|20\rangle)/\sqrt{3},\\
\end{equation*}
\begin{equation}
 \psi_{11}^{\overline{3}}=|01\rangle+e^{2\pi i/3}|12\rangle+e^{4\pi i/3}|20\rangle)/\sqrt{3},\\
\end{equation}
\begin{equation*}
 \psi_{21}^{\overline{3}}=|01\rangle+e^{4\pi i/3}|12\rangle+e^{2\pi i/3}|20\rangle)/\sqrt{3},\\
\end{equation*}
\begin{equation*}
 \psi_{02}^{\overline{3}}=(|02\rangle+|10\rangle+|21\rangle)/\sqrt{3},\\
\end{equation*}
\begin{equation*}
 \psi_{12}^{\overline{3}}=|02\rangle+e^{2\pi i/3}|10\rangle+e^{4\pi i/3}|21\rangle)/\sqrt{3},\\
\end{equation*}
\begin{equation*}
 \psi_{22}^{\overline{3}}=(|02\rangle+e^{4\pi i/3}|10\rangle+e^{2\pi i/3}|21\rangle)/\sqrt{3},\\
\end{equation*}
here, $i$, $j$ represent the subscript of system state, and $\overline{d}$ represent the dimension of system state.
The three-dimensional auxiliary state $\varphi^{\overline{3}}$ is defined as
\begin{equation*}
\varphi^{\overline{3}}=(|aa\rangle+|bb\rangle+|cc\rangle)/\sqrt{3},
\end{equation*}
where $a$, $b$, $c$ are at different DOF with $0, 1, 2$. In our scheme, nine system states can be distinguished in a hyperentangled state of the form $\psi_{ij}^{\overline{3}}\otimes\varphi^{\overline{3}}$. With the auxiliary entanglement, the hyperentangled states can be rewritten as a superposition of combinations of single photon Bell states: $\psi_{ij}^{\overline{3}}\otimes\varphi^{\overline{3}}$=$\sum\alpha^{\overline{3}}_{km}\otimes\alpha_{k'm'}^{\overline{3}}$. Here, $\alpha$ is defined as decomposition state; $k$, $m$, $k'$, $m'$ represent the subscript of decomposition state.

The three dimensional decomposition state $\alpha_{km}^{\overline{3}}$ has the form:
\begin{equation*}
 \alpha^{\overline{3}}_{00}=(|0a\rangle+|1b\rangle+|2c\rangle)/\sqrt{3},\\
\end{equation*}
\begin{equation*}
 \alpha_{10}^{\overline{3}}=(|0a\rangle+e^{2\pi i/3}|1b\rangle+e^{4\pi i/3}|2c\rangle)/\sqrt{3},\\
\end{equation*}
\begin{equation*}
 \alpha_{20}^{\overline{3}}=(|0a\rangle+e^{4\pi i/3}|1b\rangle+e^{2\pi i/3}|2c\rangle)/\sqrt{3},\\
\end{equation*}
\begin{equation*}
 \alpha_{01}^{\overline{3}}=(|0c\rangle+|1a\rangle+|2b\rangle)/\sqrt{3},\\
\end{equation*}
\begin{equation}
 \alpha_{11}^{\overline{3}}=(|0c\rangle+e^{2\pi i/3}|1a\rangle+e^{4\pi i/3}|2b\rangle)/\sqrt{3},\\
\end{equation}
\begin{equation*}
 \alpha_{21}^{\overline{3}}=(|0c\rangle+e^{4\pi i/3}|1a\rangle+e^{2\pi i/3}|2b\rangle)/\sqrt{3},\\
\end{equation*}
\begin{equation*}
 \alpha_{02}^{\overline{3}}=(|0b\rangle+|1c\rangle+|2a\rangle)/\sqrt{3},\\
\end{equation*}
\begin{equation*}
 \alpha_{12}^{\overline{3}}=(|0b\rangle+e^{2\pi i/3}|1c\rangle+e^{4\pi i/3}|2a\rangle)/\sqrt{3},\\
\end{equation*}
\begin{equation*}
 \alpha_{22}^{\overline{3}}=(|0b\rangle+e^{4\pi i/3}|1c\rangle+e^{2\pi i/3}|2a\rangle)/\sqrt{3}.\\
\end{equation*}

 In our scheme, one rule of constructing decomposition state $\alpha_{km}^{\overline{3}}$ must be mentioned. In the three-dimensional decomposition states, \\$\alpha_{k0}^{\overline{3}}=(|0\otimes a\rangle+e^{\theta i}|1\otimes b\rangle+e^{\theta i}|2\otimes c\rangle)/\sqrt{3}$, \\ $\alpha_{k1}^{\overline{3}}=(|0\otimes(a-1)\rangle+e^{\theta i}|1\otimes(b-1)\rangle+e^{\theta i}|2\otimes(c-1)\rangle)/\sqrt{3}$, \\$\alpha_{k2}^{\overline{3}}=(|0\otimes(a-2)\rangle+e^{\theta i}|1\otimes(b-2)\rangle+e^{\theta i}|2\otimes(c-2)\rangle)/\sqrt{3}$,\\
here, we take ternary operations on $a, b, c$.

 Then $\psi_{ij}^{\overline{3}}\otimes\varphi^{\overline{3}}$ can be rewritten as a superposition of nine combinations of decomposition states. We present the detailed calculation results of hyperentangled states $\psi_{00}^{\overline{3}}\otimes\varphi^{\overline{3}}$ and $\psi_{00}^{\overline{3}}\otimes\varphi^{\overline{3}}$ as an example,
\begin{equation*}
\begin{split}
 \psi_{00}^{\overline{3}}\otimes\varphi^{\overline{3}}=&\alpha_{00}^{\overline{3}}\otimes\alpha_{00}^{\overline{3}}+\alpha_{10}^{\overline{3}}\otimes\alpha_{20}^{\overline{3}}+\alpha_{20}^{\overline{3}}\otimes\alpha_{10}^{\overline{3}}\\+&\alpha_{01}^{\overline{3}}\otimes\alpha_{01}^{\overline{3}}+\alpha_{11}^{\overline{3}}\otimes\alpha_{21}^{\overline{3}}+\alpha_{21}^{\overline{3}}\otimes\alpha_{11}^{\overline{3}}\\+&\alpha_{02}^{\overline{3}}\otimes\alpha_{02}^{\overline{3}}+\alpha_{12}^{\overline{3}}\otimes\alpha_{22}^{\overline{3}}+\alpha_{22}^{\overline{3}}\otimes\alpha_{12}^{\overline{3}},
 \end{split}
 \end{equation*}
\begin{equation*}
 \begin{split}
 \psi_{10}^{\overline{3}}\otimes\varphi^{\overline{3}}=&\alpha_{00}^{\overline{3}}\otimes\alpha_{20}^{\overline{3}}+e^{2\pi i/3}\alpha_{10}^{\overline{3}}\otimes\alpha_{10}^{\overline{3}}+e^{4\pi i/3}\alpha_{20}^{\overline{3}}\otimes\alpha_{00}^{\overline{3}}\\+&\alpha_{01}^{\overline{3}}\otimes\alpha_{21}^{\overline{3}}+e^{2\pi i/3}\alpha_{11}^{\overline{3}}\otimes\alpha_{11}^{\overline{3}}+e^{4\pi i/3}\alpha_{21}^{\overline{3}}\otimes\alpha_{01}^{\overline{3}}\\+&\alpha_{02}^{\overline{3}}\otimes\alpha_{22}^{\overline{3}}+e^{2\pi i/3}\alpha_{12}^{\overline{3}}\otimes\alpha_{12}^{\overline{3}}+e^{4\pi i/3}\alpha_{22}^{\overline{3}}\otimes\alpha_{02}^{\overline{3}}.
 \end{split}
 \end{equation*}
As same form as $ \psi_{00}^{\overline{3}}\otimes\varphi^{\overline{3}} $ and $ \psi_{10}^{\overline{3}}\otimes\varphi^{\overline{3}} $, each three-dimensional hyperentangled state can be rewritten as a unique superposition of nine of 81 possible combinations of decomposition states. Therefore, the nine three-dimensional hyperentangled states can be distinguished according to their unique decomposition form.
We present the decomposition form of nine three-dimensional hyperentangled states in the TABLE I.

\begin{table}[!h]
\begin{center}
\centering
\caption{\textbf{decomposition form of three-dimensional hyperentangled states.} The final row list the decomposition form of nine hyperentangled states.}
\renewcommand{\arraystretch}{1.3}
\begin{tabular}{c|m{5.2cm}<{\centering}}
\hline
\hline
Hyperentangled state & decomposition form\\
\hline
 $\psi_{00}^{\overline{3}}\otimes\varphi^{\overline{3}}$ &$\alpha_{00}^{\overline{3}}\otimes\alpha_{00}^{\overline{3}}, $ $\alpha_{10}^{\overline{3}}\otimes\alpha_{20}^{\overline{3}}, $ $\alpha_{20}^{\overline{3}}\otimes\alpha_{10}^{\overline{3}} $ $\alpha_{01}^{\overline{3}}\otimes\alpha_{01}^{\overline{3}}, $ $\alpha_{11}^{\overline{3}}\otimes\alpha_{21}^{\overline{3}}, $ $\alpha_{21}^{\overline{3}}\otimes\alpha_{11}^{\overline{3}} $ $\alpha_{02}^{\overline{3}}\otimes\alpha_{02}^{\overline{3}}, $ $\alpha_{12}^{\overline{3}}\otimes\alpha_{22}^{\overline{3}}, $ $\alpha_{22}^{\overline{3}}\otimes\alpha_{12}^{\overline{3}}$\\
\hline
$\psi_{10}^{\overline{3}}\otimes\varphi^{\overline{3}}$ &$\alpha_{00}^{\overline{3}}\otimes\alpha_{20}^{\overline{3}}, $ $\alpha_{10}^{\overline{3}}\otimes\alpha_{10}^{\overline{3}}, $ $\alpha_{20}^{\overline{3}}\otimes\alpha_{00}^{\overline{3}} $ $\alpha_{01}^{\overline{3}}\otimes\alpha_{21}^{\overline{3}}, $ $\alpha_{11}^{\overline{3}}\otimes\alpha_{11}^{\overline{3}}, $ $\alpha_{21}^{\overline{3}}\otimes\alpha_{01}^{\overline{3}} $ $\alpha_{02}^{\overline{3}}\otimes\alpha_{22}^{\overline{3}}, $ $\alpha_{12}^{\overline{3}}\otimes\alpha_{12}^{\overline{3}}, $ $\alpha_{22}^{\overline{3}}\otimes\alpha_{02}^{\overline{3}}$\\
\hline
$\psi_{20}^{\overline{3}}\otimes\varphi^{\overline{3}}$ &$\alpha_{00}^{\overline{3}}\otimes\alpha_{10}^{\overline{3}}, $ $\alpha_{10}^{\overline{3}}\otimes\alpha_{00}^{\overline{3}}, $ $\alpha_{20}^{\overline{3}}\otimes\alpha_{20}^{\overline{3}} $ $\alpha_{01}^{\overline{3}}\otimes\alpha_{11}^{\overline{3}}, $ $\alpha_{11}^{\overline{3}}\otimes\alpha_{01}^{\overline{3}}, $ $\alpha_{21}^{\overline{3}}\otimes\alpha_{21}^{\overline{3}} $ $\alpha_{02}^{\overline{3}}\otimes\alpha_{12}^{\overline{3}}, $ $\alpha_{12}^{\overline{3}}\otimes\alpha_{02}^{\overline{3}}, $ $\alpha_{22}^{\overline{3}}\otimes\alpha_{22}^{\overline{3}}$\\
\hline
$\psi_{01}^{\overline{3}}\otimes\varphi^{\overline{3}}$& $\alpha_{00}^{\overline{3}}\otimes\alpha_{01}^{\overline{3}}, $ $\alpha_{10}^{\overline{3}}\otimes\alpha_{21}^{\overline{3}}, $ $\alpha_{20}^{\overline{3}}\otimes\alpha_{11}^{\overline{3}} $ $\alpha_{01}^{\overline{3}}\otimes\alpha_{02}^{\overline{3}}, $ $\alpha_{11}^{\overline{3}}\otimes\alpha_{22}^{\overline{3}}, $ $\alpha_{21}^{\overline{3}}\otimes\alpha_{12}^{\overline{3}} $ $\alpha_{21}^{\overline{3}}\otimes\alpha_{00}^{\overline{3}}, $ $\alpha_{12}^{\overline{3}}\otimes\alpha_{20}^{\overline{3}}, $ $\alpha_{21}^{\overline{3}}\otimes\alpha_{00}^{\overline{3}}$\\
\hline
$\psi_{11}^{\overline{3}}\otimes\varphi^{\overline{3}}$ &$\alpha_{00}^{\overline{3}}\otimes\alpha_{21}^{\overline{3}}, $ $\alpha_{10}^{\overline{3}}\otimes\alpha_{11}^{\overline{3}}, $ $\alpha_{20}^{\overline{3}}\otimes\alpha_{01}^{\overline{3}} $ $\alpha_{01}^{\overline{3}}\otimes\alpha_{22}^{\overline{3}}, $ $\alpha_{11}^{\overline{3}}\otimes\alpha_{12}^{\overline{3}}, $ $\alpha_{21}^{\overline{3}}\otimes\alpha_{02}^{\overline{3}} $ $\alpha_{02}^{\overline{3}}\otimes\alpha_{20}^{\overline{3}}, $ $\alpha_{12}^{\overline{3}}\otimes\alpha_{10}^{\overline{3}}, $ $\alpha_{22}^{\overline{3}}\otimes\alpha_{00}^{\overline{3}}$\\
\hline
$\psi_{21}^{\overline{3}}\otimes\varphi^{\overline{3}}$ &$\alpha_{00}^{\overline{3}}\otimes\alpha_{11}^{\overline{3}}, $ $\alpha_{10}^{\overline{3}}\otimes\alpha_{01}^{\overline{3}}, $ $\alpha_{20}^{\overline{3}}\otimes\alpha_{21}^{\overline{3}} $ $\alpha_{01}^{\overline{3}}\otimes\alpha_{12}^{\overline{3}}, $ $\alpha_{11}^{\overline{3}}\otimes\alpha_{02}^{\overline{3}}, $ $\alpha_{21}^{\overline{3}}\otimes\alpha_{22}^{\overline{3}} $ $\alpha_{02}^{\overline{3}}\otimes\alpha_{10}^{\overline{3}}, $ $\alpha_{12}^{\overline{3}}\otimes\alpha_{00}^{\overline{3}}, $ $\alpha_{22}^{\overline{3}}\otimes\alpha_{20}^{\overline{3}}$\\
\hline
$\psi_{02}^{\overline{3}}\otimes\varphi^{\overline{3}}$ &$\alpha_{00}^{\overline{3}}\otimes\alpha_{02}^{\overline{3}}, $ $\alpha_{10}^{\overline{3}}\otimes\alpha_{22}^{\overline{3}}, $ $\alpha_{20}^{\overline{3}}\otimes\alpha_{12}^{\overline{3}} $ $\alpha_{01}^{\overline{3}}\otimes\alpha_{00}^{\overline{3}}, $ $\alpha_{11}^{\overline{3}}\otimes\alpha_{20}^{\overline{3}}, $ $\alpha_{21}^{\overline{3}}\otimes\alpha_{10}^{\overline{3}} $ $\alpha_{02}^{\overline{3}}\otimes\alpha_{01}^{\overline{3}}, $ $\alpha_{12}^{\overline{3}}\otimes\alpha_{21}^{\overline{3}}, $ $\alpha_{22}^{\overline{3}}\otimes\alpha_{11}^{\overline{3}}$\\
\hline
$\psi_{12}^{\overline{3}}\otimes\varphi^{\overline{3}}$ &$\alpha_{00}^{\overline{3}}\otimes\alpha_{22}^{\overline{3}}, $ $\alpha_{10}^{\overline{3}}\otimes\alpha_{12}^{\overline{3}}, $ $\alpha_{20}^{\overline{3}}\otimes\alpha_{02}^{\overline{3}} $ $\alpha_{01}^{\overline{3}}\otimes\alpha_{20}^{\overline{3}}, $ $\alpha_{11}^{\overline{3}}\otimes\alpha_{10}^{\overline{3}}, $ $\alpha_{21}^{\overline{3}}\otimes\alpha_{00}^{\overline{3}} $ $\alpha_{02}^{\overline{3}}\otimes\alpha_{21}^{\overline{3}}, $ $\alpha_{12}^{\overline{3}}\otimes\alpha_{11}^{\overline{3}}, $ $\alpha_{22}^{\overline{3}}\otimes\alpha_{01}^{\overline{3}}$\\
\hline
$\psi_{22}^{\overline{3}}\otimes\varphi^{\overline{3}}$ &$\alpha_{00}^{\overline{3}}\otimes\alpha_{12}^{\overline{3}}, $ $\alpha_{10}^{\overline{3}}\otimes\alpha_{02}^{\overline{3}}, $ $\alpha_{20}^{\overline{3}}\otimes\alpha_{22}^{\overline{3}} $ $\alpha_{01}^{\overline{3}}\otimes\alpha_{10}^{\overline{3}}, $ $\alpha_{11}^{\overline{3}}\otimes\alpha_{00}^{\overline{3}}, $ $\alpha_{21}^{\overline{3}}\otimes\alpha_{20}^{\overline{3}} $ $\alpha_{02}^{\overline{3}}\otimes\alpha_{11}^{\overline{3}}, $ $\alpha_{12}^{\overline{3}}\otimes\alpha_{21}^{\overline{3}}, $ $\alpha_{22}^{\overline{3}}\otimes\alpha_{21}^{\overline{3}}$\\
\hline
\hline
\bottomrule
\end{tabular}
\end{center}
\end{table}

The TABLE I can be summarized as an equation
\begin{equation}
 \psi_{ij}^{\overline{3}}\otimes\varphi^{\overline{3}}=\sum_{k,m=0}^{2,2}e^{\theta i}*\alpha^{\overline{3}}_{km}\otimes\alpha^{\overline{3}}_{(2k+2i)mod3,(m+j)mod3},
\end{equation}

\begin{figure*}[tbph]
\includegraphics [width= 1\textwidth]{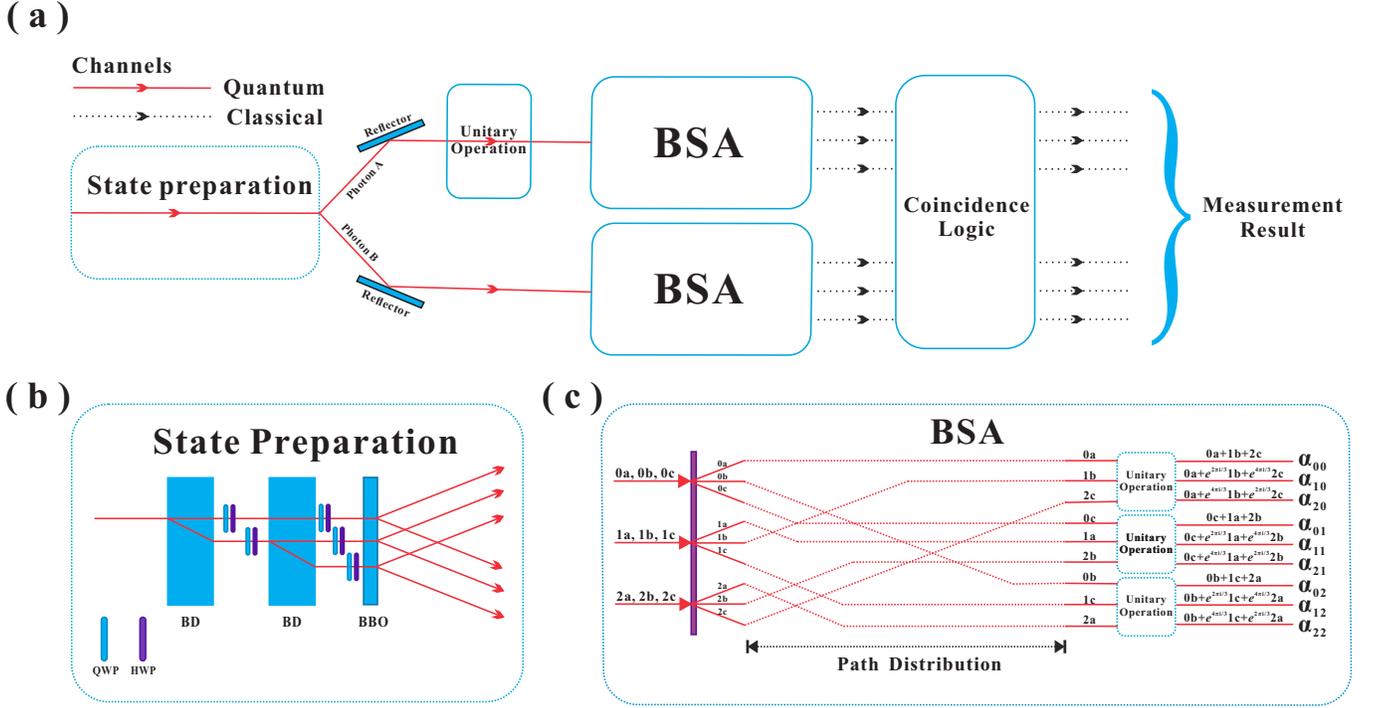}
\caption{ \textbf{Experimental scheme.} \textbf{(a)} Whole process of the experimental scheme. Firstly, we prepare an initial system state and auxiliary state $\psi_{00}^{\overline{3}}\otimes\varphi^{3}$. Then we complete the preparation of nine Bell states through three-dimensional arbitrary unitary operation of path DOF. Finally, we distinguish the decomposition state $\alpha{ij}^{\overline{3}}$ via Bell state analyser (BSA), and complete the HDBSM of the system state by coincidence.
 \textbf{(b)}  State preparation scheme. Photons are created via spontaneous parametric down conversion (SPDC) by pumping type-I phase matched $\beta$-barium borate (BBO) nonlinear crystals with a CW laser. Because SPDC satisfying OAM conservation, thus the state prepared by SPDC comes to be $(|00\rangle+|11\rangle+|22\rangle)\otimes(|aa\rangle+|bb\rangle+|cc\rangle)/3$. Through the unitary operation on one photon's path DOF, we can change the system state into 9 Bell states. \textbf{(c)} Path-OAM BSA. It is the realization of three-dimensional BSM for single-photon with different DOFs (path and OAM). We convert the OAM DOF into path DOF by splitting the beam according to its OAM, and then the BSM between two DOFs can be completed by appropriate path distribution and unitary operation.
 }
\end{figure*}

here, $i$, $j$ represent the subscript of three-dimensional system state; $\alpha^{\overline{3}}_{km}$ represent the decomposition states on Bob side; $\alpha^{\overline{3}}_{(2k+2i)mod3,(m+j)mod3}$ represent the decomposition states on the Alice side; $\theta$ represent the phase of combinations of decomposition states. Each three-dimensional hyperentangled state corresponds to a unique equation, thus we can complete the three-dimensional BSM according to Eq. (3). Then we extend Eq. (3) used in three-dimensional BSM to the BSM in higher dimensions.

We verified the two-dimensional BSM \cite{two-dimension1} in our scheme, result shows that two-dimensional BSM can be perfect implemented in our scheme. The equation of two-dimensional BSM is:
\begin{equation}
 \psi_{ij}^{\overline{2}}\otimes\varphi^{\overline{2}}=\sum_{k,m=0}^{1,1}e^{\theta i}*\alpha_{k m}^{\overline{2}}\otimes\alpha^{\overline{2}}_{(k+i)mod2,(m+j)mod2}
\end{equation}

For four-dimensional BSM, the measurement scheme is as same as three-dimensional BSM. Each hyperentangled state is a unique superposition of 16 of the 256 possible combinations of decomposition states. For example, the decomposition form of $\psi_{23}^{\overline{4}}\otimes\varphi^{\overline{4}}$ are
\begin{equation*}
 \begin{array}{cccc}

 \alpha_{00}^{\overline{4}}\otimes\alpha_{23}^{\overline{4}}, &\alpha_{10}^{\overline{4}}\otimes\alpha_{13}^{\overline{4}}, &\alpha_{20}^{\overline{4}}\otimes\alpha_{03}^{\overline{4}}, &\alpha_{30}^{\overline{4}}\otimes\alpha_{33}^{\overline{4}}, \\
 \alpha_{01}^{\overline{4}}\otimes\alpha_{20}^{\overline{4}}, &\alpha_{11}^{\overline{4}}\otimes\alpha_{10}^{\overline{4}}, &\alpha_{21}^{\overline{4}}\otimes\alpha_{00}^{\overline{4}}, &\alpha_{31}^{\overline{4}}\otimes\alpha_{30}^{\overline{4}}, \\
 \alpha_{02}^{\overline{4}}\otimes\alpha_{21}^{\overline{4}}, &\alpha_{12}^{\overline{4}}\otimes\alpha_{11}^{\overline{4}}, &\alpha_{22}^{\overline{4}}\otimes\alpha_{01}^{\overline{4}}, &\alpha_{32}^{\overline{4}}\otimes\alpha_{31}^{\overline{4}}, \\
 \alpha_{03}^{\overline{4}}\otimes\alpha_{22}^{\overline{4}}, &\alpha_{13}^{\overline{4}}\otimes\alpha_{12}^{\overline{4}}, &\alpha_{23}^{\overline{4}}\otimes\alpha_{02}^{\overline{4}}, &\alpha_{33}^{\overline{4}}\otimes\alpha_{32}^{\overline{4}}. \\
 \end{array}
\end{equation*}
Same as three-dimensional hyperentangled states, four-dimensional hyperentangled states can also be rewritten as:
\begin{equation}
 \psi_{ij}^{\overline{4}}\otimes\varphi^{\overline{4}}=\sum_{k,m=0}^{3,3}e^{\theta i}*\alpha_{k m}^{\overline{4}}\otimes\alpha^{\overline{4}}_{(3k+3i)mod4,(m+j)mod4}.
\end{equation}

The construction rule of four-dimensional decomposition states is same as three-dimensional decomposition states, \\$\alpha_{k0}^{\overline{4}}=(|0\otimes a\rangle+e^{\theta i}|1\otimes b\rangle+e^{\theta i}|2\otimes c\rangle+e^{\theta i}|3\otimes d\rangle)/2$, \\
$\alpha_{k1}^{\overline{4}}=(|0\otimes(a-1)\rangle+e^{\theta i}|1\otimes(b-1)\rangle+e^{\theta i}|2\otimes(c-1)\rangle+e^{\theta i}|3\otimes(d-1)\rangle)/2$, \\
$\alpha_{k2}^{\overline{4}}=(|0\otimes(a-2)\rangle+e^{\theta i}|1\otimes(b-2)\rangle+e^{\theta i}|2\otimes(c-2)\rangle+e^{\theta i}|3\otimes(d-2)\rangle)/2$, \\
$\alpha_{k1}^{\overline{4}}=(|0\otimes(a-3)\rangle+e^{\theta i}|1\otimes(b-3)\rangle+e^{\theta i}|2\otimes(c-3)\rangle+e^{\theta i}|3\otimes(d-3)\rangle)/2$, \\
here, we take quaternary operations on $a, b, c, d$.

We can follow the three-dimensional BSM scheme, achieve the complete four-dimensional BSM according to Eq. (5).
The scheme of three-dimensional BSM can also be extended to arbitrary high-dimensional BSM, thus we derive the equation suitable for arbitrary dimension:
\begin{equation}
\begin{split}
 &\psi_{ij}^{\overline{d}}\otimes\varphi^{\overline{d}}= \\
 &\sum_{k,m=0}^{d-1,d-1}e^{\theta i}*\alpha_{k m}^{\overline{d}}\otimes\alpha^{\overline{d}}_{((d-1)k+(d-1)i)mod (d),(m+j)mod (d)}.
\end{split}
\end{equation}
The unique superposition of $d\times d$ combinations of arbitrary high-dimensional decomposition states can be directly listed from Eq. (6), thus we can realize the arbitrary high-dimensional BSM.

We have also designed an experimental scheme for three-dimensional BSM. In our experimental scheme, the decomposition state $\alpha_{km}^{\overline{3}}$ is composed of path \cite{path} and OAM \cite{OAM} DOFs. We use path as system DOF and OAM as auxiliary DOF, complete the three-dimensional BSM via linear optics. In the HDBSM process, three parts are needed: preparation of system state and auxiliary state, unitary operation of single particle with the same DOF, and path-OAM BSM.

As shown in Fig. 1, firstly, we implement the system state and auxiliary state preparation via a standard three-dimensional path entanglement preparation. System state $\psi_{00}^{\overline{3}}=(|00\rangle+|11\rangle+|22\rangle)/\sqrt{3}$ \cite{qutrits} can be prepared by spontaneous parametric down-conversion (SPDC). At the same time, because of the OAM conservation in the SPDC process, there are OAM entanglements in each path. Consequently, the three-dimensional hyperentangled state $(|00\rangle+|11\rangle+|22\rangle)\otimes(|aa\rangle+|bb\rangle+|cc\rangle)/3$ is generated. Next, we construct an unitary operation \cite{unitary operation} of three-dimensional path DOF. Nine system states $\psi_{ij}^{\overline{3}}$ can be created by unitary operating on input state$(|00\rangle+|11\rangle+|22\rangle)/\sqrt{3}$. Finally, we use our designed HDBSM to distinguish nine three-dimensional hyperentangled states. In fig. 1 (c), each beam was splitted into three beams according to the OAM (-1, 0, +1, where -1 encodes as $a$; 0 encodes as $b$; +1 encodes as $c$). In this way, the OAM DOF is converted into the path DOF. Thus we can distinguish all the nine three-dimensional hyperentangled states as shown in Eq. (2).

In detail, $\alpha_{00}^{\overline{3}}$, $\alpha_{10}^{\overline{3}}$ and $\alpha_{20}^{\overline{3}}$ are composed of $|0a\rangle$, $|1b\rangle$, $|2c\rangle$, so we distribute $|0a\rangle$, $|1b\rangle$, $|2c\rangle$ into one three-dimensional unitary operation, thus via unitary operation, $\alpha_{00}^{\overline{3}}$, $\alpha_{10}^{\overline{3}}$ and $\alpha_{20}^{\overline{3}}$ are projected to three discrete paths. The procedure can be represented by the unitary operations:

\begin{equation*}
\frac{1}{3}{
\left( \begin{array}{ccc}
1 & 1 & 1\\
1 & e^{-2\pi i/3} & e^{-4\pi i/3}\\
1 & e^{-4\pi i/3} & e^{-2\pi i/3}
\end{array}
\right)}
\cdot
\left(
 \begin{array}{c}
  1 \\
  1 \\
  1
 \end{array}
 \right)
=
\left(
 \begin{array}{c}
  1 \\
  0 \\
  0
 \end{array}
 \right),\\
\end{equation*}
\begin{equation*}
\frac{1}{3}{
\left( \begin{array}{ccc}
1 & 1 & 1\\
1 & e^{-2\pi i/3} & e^{-4\pi i/3}\\
1 & e^{-4\pi i/3} & e^{-2\pi i/3}
\end{array}
\right)}
\cdot
\left(
 \begin{array}{c}
  1 \\
  e^{2\pi i/3} \\
  e^{4\pi i/3}
 \end{array}
 \right)
=
\left(
 \begin{array}{c}
  0 \\
  1 \\
  0
 \end{array}
 \right),\\
\end{equation*}

\begin{equation*}
\frac{1}{3}{
\left( \begin{array}{ccc}
1 & 1 & 1\\
1 & e^{-2\pi i/3} & e^{-4\pi i/3}\\
1 & e^{-4\pi i/3} & e^{-2\pi i/3}
\end{array}
\right)}
\cdot
\left(
 \begin{array}{c}
  1 \\
  e^{4\pi i/3} \\
  e^{2\pi i/3}
 \end{array}
 \right)
=
\left(
 \begin{array}{c}
  0 \\
  0 \\
  1
 \end{array}
 \right),\\
\end{equation*}
here, we use the column vector to represent the decomposition state.
As the unitary operation result shown, $\alpha_{00}^{\overline{3}}$ only exports from the first path, $\alpha_{10}^{\overline{3}}$ only exports from the second path, and $\alpha_{20}^{\overline{3}}$ only exports from the third path. Same as the above unitary operation, $\alpha_{01}^{\overline{3}}$, $\alpha_{11}^{\overline{3}}$, $\alpha_{21}^{\overline{3}}$, $\alpha_{02}^{\overline{3}}$, $\alpha_{12}^{\overline{3}}$ and $\alpha_{22}^{\overline{3}}$ can also be projected to discrete paths via unitary operation. According to Table I, one can complete HDBSM for system state by coincidence measurement.

Arbitrary HDBSM can be realized by extending the three-dimensional BSM scheme. In the d-dimensional BSM, the d-dimensional OAM auxiliary entanglement and the d-dimensional unitary operation are needed. The construction method of d-dimensional decomposition states can refer the construction of three-dimensional decomposition states.

In conclusion, we propose a method for recognizing high-dimensional Bell states. By using two degrees of freedom of the same particle, we can complete the HDBSM only by BSM between different degrees of freedom in the local region. We deduce the recognition method of arbitrary dimension entangled Bell states, and we design a three-dimensional BSM experimental scheme. The scheme will greatly promote the development of high-dimensional quantum information tasks, such as high-dimensional dense coding, high-dimensional swap, high-dimensional teleportation.

This work was supported by the National Key Research and Development Program of China (No.\ 2017YFA0304100, No. 2016YFA0301300 and No. 2016YFA0301700), NSFC (Nos. 11774335, 61327901, 11474268, 11504253, 11874345, 11821404), the Key Research Program of Frontier Sciences, CAS (No.\ QYZDY-SSW-SLH003), the Fundamental Research Funds for the Central Universities, and Anhui Initiative in Quantum Information Technologies (Nos.\ AHY020100, AHY060300).

\end{document}